\documentclass[epj]{svjour}
\usepackage{graphicx}
\usepackage[usenames,dvipsnames]{xcolor}
\usepackage{subfig}
\usepackage{pifont}

\usepackage{latexsym}

\begin{document}
\title{Distinguishing noisy crystalline structures using bond orientational order parameters}
\author{Jan Haeberle\inst{1} \and Matthias Sperl\inst{1,2} \and Philip Born\inst{1}
}                     
%
%
\institute{Institut f\"{u}r Materialphysik im Weltraum, Deutsches Zentrum f\"{u}r Luft-und Raumfahrt, 51170 K\"{o}ln, Germany \and Institut f\"ur Theoretische Physik, Universit\"at zu K\"oln, 50937 K\"oln, Germany}
\date{Received: date / Revised version: date}
%
\abstract{
The bond orientational order parameters originally introduced by Steinhardt \emph{et. al.} [Phys. Rev. B \textbf{28}, 784 (1983)] are a common tool for local structure characterization in soft matter studies. Recently, Mickel \emph{et. al.} [J. Chem. Phys. \textbf{138}, 044501 (2013)] highlighted problems of the bond orientational order parameters due to the ambiguity of the underlying neighbourhood definition. Here we show the difficulties of distinguish common structures like FCC- and BCC-based structures with the suggested neighbourhood definitions when noise is introduced. We propose a simple improvement to the neighbourhood definition that results in robust and continuous bond orientational order parameters with which we can accurately distinguish crystal structures even when noise is present.
\PACS{
      {PACS-key}{discribing text of that key}   \and
      {PACS-key}{discribing text of that key}
     } 
} 
\maketitle


\section{\label{sec:intro} Introduction}

A great benefit of studies using soft matter systems is observability on the local scale. Thermodynamic and statistical processes like phase transitions and crystallization can be studied on the level of individual constituents, like solid particles in colloids, dusty plasma and granular media, gas bubbles in foams or droplets in emulsion \cite{allahyarov_crystallization_2015,gasser_real-space_2001,chu_direct_1994,arp_dust_2004,nefedov_pke-nefedov:_2003,panaitescu_epitaxial_2014,aste_investigating_2004}. 

Steinhardt and co-workers introduced the bond orientational order parameters $q_l$ \cite{Steinhardt_bond-orientational_1983} as a useful local order metric in such studies. The arrangement structure of an individual constituent is quantified by the angles among its 'bonds', i.e. the directions to the neighbouring constituents (see Sec.~\ref{sec:state} for details). The $q_l$ are rotation invariant and are sensitive to symmetries in the bond angles, thus comparing measured values to values of ideal lattices allows for identifying crystalline arrangements on the level of individual constituents.

These Steinhardt bond orientational order parameters are widely used \cite{lechner_accurate_2008,mickel_shortcomings_2013,schroder-turk_minkowski_2013}, but are not uniquely defined due to the ambiguity of neighborhood. Here we continue a discussion by Mickel \textit{et al.} on different definitions and quantification of neighbourhood \cite{mickel_shortcomings_2013}, and show an optimized definition in the presence of noise. Noise implies that the measured positions of the constituents fluctuate around the equilibrium, mean or real positions. Such noise may arise from a finite resolution of the experimental technique and the resulting uncertainty in measured positions. Intrinsic noise also emerges from thermal fluctuations of positions and from polydispersity of constituent shape and interactions \cite{gasser_real-space_2001,pusey_effect_1987}. The measured arrangement thus will usually exhibit a distribution of local arrangements of the constituents. The local order metric reflects this distribution and extraction of information on underlying crystal structures thus needs additional measures. For example, the analysis was constricted to statistical decomposition of the measured distributions of the $q_l$ into the distributions of $q_l$ obtained numerically from different noisy crystal structures in order to retrieve information on the apparent fractions of the respective crystal \cite{gasser_real-space_2001,ten_wolde_numerical_1995,rein_ten_wolde_numerical_1996,volkov_molecular_2002}. Assignment of crystal structures to $q_l$ of individual constituents in studies with noise relies on the distance of the individual $q_l$ to the $q_l$ of the ideal crystal lattice site. Sensitivity of this approach was improved by using more than one $q_l$ and mapping regions in the $q_l-q_m$-plane to certain crystal structures or by using information on the neighbours of the neighbours \cite{lechner_accurate_2008,desgranges_crystallization_2008,rietz_nucleation_2018}.

However, the bond orientational order parameter has to rely on a neighbourhood definition. This quantification of neighbourhood itself is affected by the noise to a degree which depends on the used definition, as we highlight in Sec.~\ref{sec:resultsI}. Consequently, measured distributions of $q_l$ may not reflect the distribution of local arrangements and structure identification turns ambiguous. Mickel \textit{et al.} have discussed different neighbourhood definitions and proposed an unambiguous and continuous morphometric neighbourhood definition (see Sec.~\ref{sec:state}) \cite{mickel_shortcomings_2013}. We suggest a simple optimization of the morphometric neighbourhood definition. This suggested definition minimizes detrimental effects of noise and facilitates discrimination between the common crystal structures like body-centered cubic (BCC), face-centered cubic (FCC) or hexagonally close-packed (HCP) lattices even in the presence of noise.

We describe in Sec.~\ref{sec:methods} how we generate the test data. In Sec.~\ref{sec:state} the Steinhardt bond orientational parameters and the modifications introduced by  Mickel \textit{et al.} are discussed. We also motivate our optimization in this section. The effects of noise and the difficulty of discriminating BCC-, FCC- and HCP-based structures are highlighted in Sec.~\ref{sec:resultsI}. We show that the improved version of the morphometric neighbourhood definition enables the desired structure differentiation among the common crystal structures with noise.


\section{\label{sec:methods} Generation of structures and tessellation }

\begin{figure}
\centering
\includegraphics[width=\linewidth]{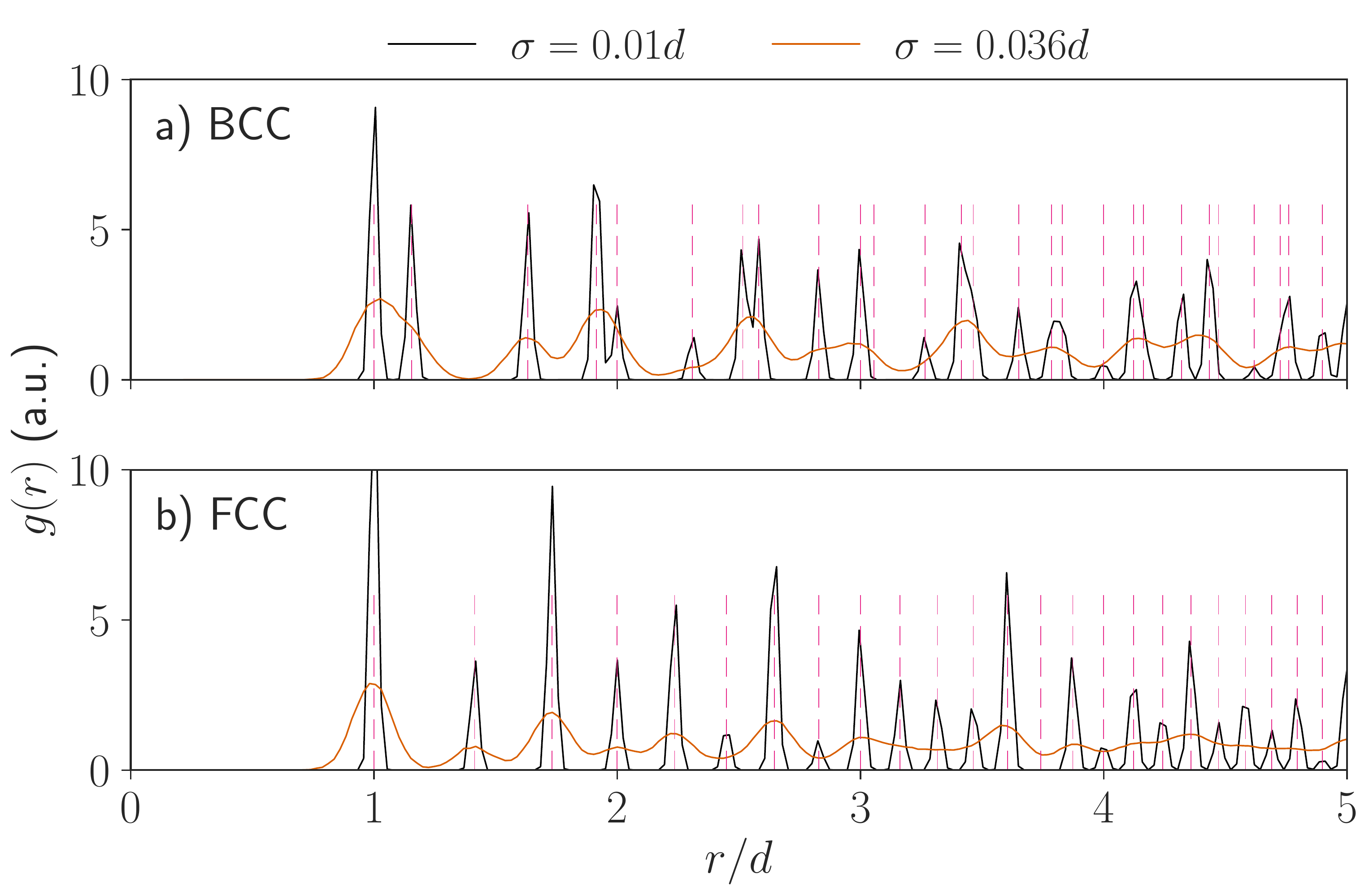}
\caption{Radial distribution functions $g(r)$ of a) BCC-based and b) FCC-bsed structures with noise of $\sigma=0.01d$ and $\sigma=0.036d$ as they are used in the analysis (see Sec.~\ref{sec:methods} for details of generating the structures). The distance $r$ is scaled by the lattice constant $d$. The dashed lines indicate the peak positions of the noise-free ideal lattice. The structures with noise still exhibit long-ranged positional correlations and distinguishable local environments.}
\label{fig:gr}
\end{figure}

We numerically mimic the effect of the various noise mechanisms mentioned in Sec.~\ref{sec:intro} in the positional data. We start by creating ideal HCP, FCC and BCC lattices with $\approx4500$ lattice sites by placing a site at integer multiples of the primitive vectors for the corresponding crystal structure.
We then apply normally distributed noise to every Cartesian component of each lattice site to simulate different degrees of noise. The normal distributions from which the x-, y- and z-displacements are drawn have a mean of 0 and standard deviations from $\sigma=0.01d$ up to $\sigma=0.1d$ in 10 logarithmically spaced steps, where $d$ is the lattice constant. We note that normally distributed noise applied to each component results in a Maxwell-Boltzmann distribution of the absolute radial displacements.
The root mean square displacement is related to the component-wise standard deviation via $$\sqrt{\langle r^2 \rangle}=\sqrt{3}\sigma.$$
An common criterion for predicting melting of a crystal is the Lindemann criterion \cite{lindemann_uber_1910}, which states that thermal motion in a crystal leads to melting once the root mean square displacement exceeds a certain fraction $\delta_l$ of the lattice constant. Typical values of this critical fraction are $\delta_l\approx 0.22$ \cite{cahn_melting_2001}, which is still higher than the highest fraction we tested of $\sqrt{\langle r^2 \rangle} / d = \sqrt{3}\cdot0.1 \approx0.171$. Therefore, crystals can realistically exhibit deviations as high as the ones we investigate here.

Our protocol leads to 10 homogeneous structures with increasing levels of noise. Figure~\ref{fig:gr} shows the radial distribution functions of the BCC- and FCC-based structures with noise with $\sigma=0.01d$ and $\sigma=0.036d$. The structures exhibit long-ranged positional correlations and distinguishable local environments even with noise, only the correlation peaks become increasingly broadened by the random displacements.

We use the Voro++-code for a Voronoi tesselation of the test structures \cite{rycroft_voro++:_2009}. The code provides vertices, edges and area of the facets of the Voronoi cells of the sites. We include all sites for the calculation of the bond orientational order but we exclude in our results the boundary sites, i.e. those that share a facet with with the lattice boundary, and any site sharing facet with a boundary site. This exclusion of two layers of sites is necessary when working with BCC-based structures, since the Voronoi facets of the sites are determined both by the nearest and the next-nearest neighbours (see Fig.~\ref{fig:qlideal}. The data shown is thus for about 3400 sites each.

These test structures are consequently evaluated using the Steinhardt bond orientational order parameter using different neighbourhood definitions.


\section{\label{sec:state} The order metric and the neighbourhood definition}

\begin{table}
\centering
\begin{tabular}{c|c|c|c|c}
Structure & neighbourhood def. & $q4$ & $q6$ & Symbol\\\hline
HCP &Thr., $\alpha=0,1,2$& 0.097  & 0.48 & $\triangleleft$\\
FCC &Thr., $\alpha=0,1,2$& 0.19 & 0.57 & \ding{73}\\
BCCt &Threshold& 0.51 & 0.63 & $\Diamond$\\
BCC0 &$\alpha$=0& 0.036 & 0.51 & $\triangleright$\\
BCC1 &$\alpha$=1& 0.22& 0.57 & $\Box$\\
BCC2 &$\alpha$=2& 0.38 & 0.60 & \ding{109}
\end{tabular}
\caption{$q_4$ and $q_6$ values for noise-free hexagonal close-packed (HCP), face-centered cubic (FCC) and body-centered cubic (BCC) lattices. The $q_l$-values of the BCC lattice sites depend sensitively on the definition of neighbourhood, in contrast to the values for FCC and HCP lattice sites. $q_l$ values are given for neighbourhood definitions using a fixed threshold of 1.1$d$ and using facet-weighting $\alpha$ of order 0, 1, and 2.}
\label{tab:ql}
\end{table}

\begin{figure}
\includegraphics[width=\linewidth]{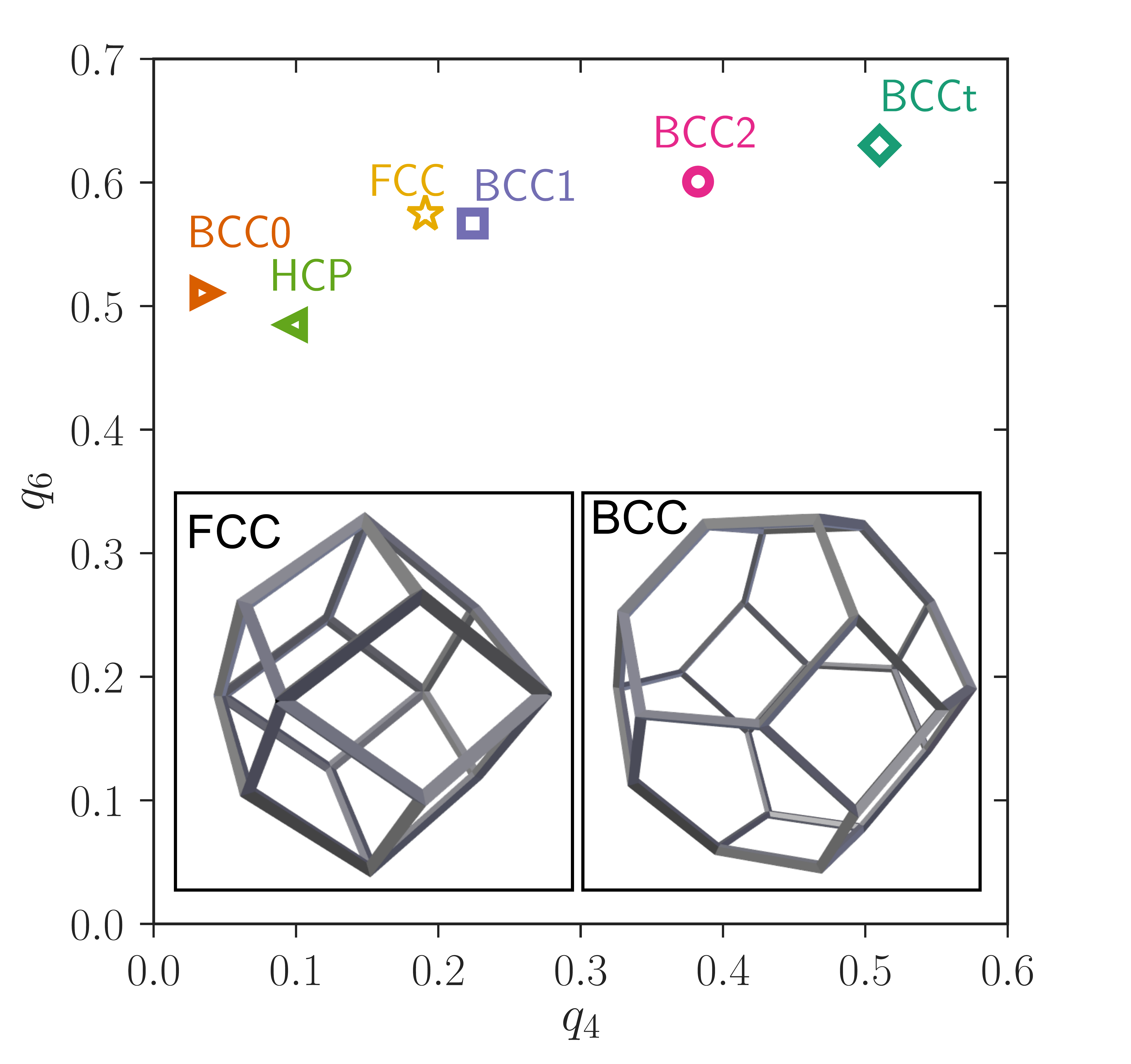}
\caption{The bond orientational order of the ideal noise-free lattice sites for the four different neighbourhood definitions, given as position in the $q_4$-$q_6$ plane (compare Tab.~\ref{tab:ql}). The position of the BCC lattice site shifts closer or further from the close-packed lattice sites, depending on whether a fixed threshold ("t") or a facet-weighting $\alpha$ of order 0, 1, and 2 is used in the neighbourhood definition. The inset shows the outline of the Voronoi cell of an FCC (left) and BCC (right) lattice site.}
\label{fig:qlideal}
\end{figure}

The Steinhardt bond orientational parameters $q_l$ are rotation invariant metrics which are sensitive to periodicity in the orientations between neighbouring sites \cite{Steinhardt_bond-orientational_1983}. The $q_l$ are calculated for any site $i$ in the medium according to:
\begin{equation}
q_l^{(i)}=\sqrt{\frac{4\pi}{2l+1}\sum_{m=-l}^l\left|\frac{1}{Z_i}\sum_{j=1}^{Z_i}Y_l^m[\theta(\mathbf{x_j}), \phi(\mathbf{x_j})]\right|^2},
\label{eq:boo}
\end{equation}
where $Z_i$ is the number of neighbours of the test site $i$, $Y_l^m$ are the spherical harmonics functions of degree $l$ and order $m$, $\mathbf{x_j}$ is the vector connecting the test site with its $j$th neighbour and $\theta$ and $\phi$ are the angles of $\mathbf{x_j}$ in spherical coordinates. 

An obvious method to determine the neighbours is to count all site positions within a cutoff distance $r_c$ as neighbours \cite{allahyarov_crystallization_2015,gasser_real-space_2001,rein_ten_wolde_numerical_1996,volkov_molecular_2002}. This approach is easy to implement, but introduces the parameter $r_c$. This parameter is either set arbitrarily \cite{allahyarov_crystallization_2015,rein_ten_wolde_numerical_1996} or is defined by the first minimum of the radial distribution function $g(r)$ \cite{gasser_real-space_2001,volkov_molecular_2002}. A problem with this definition of $r_c$ is apparent from Fig.~\ref{fig:gr}. The first minimum of $g(r)$ may readily be at distances smaller or larger than the shell of the next nearest neighbours in the BCC structure, depending on the level of noise. The usage of a fixed $r_c$ to determine neighbourhood also implies a discontinuous quantification, as infinitesimal shifts in position can produce or annihilate a bond to a neighbor.

A parameter-free determination of neighbourhood is based on the Delaunay-triangulation. The De\-lau\-nay-\-tri\-an\-gu\-la\-tion creates the dual graph of the Voronoi-tesselation, thus every site that shares a facet of the Voronoi cell of the test site is counted as a neighbour \cite{fortune_voronoi_1995}. As Mickel \textit{et al.} pointed out, this method of triangulation is parameter-free, but still not continuous \cite{mickel_shortcomings_2013}. Infinitesimal changes in the site positions can lead to forming or disappearance of facets and thus of bonds to neighbours, resulting in a discontinuous change in the value of $q_l$. They instead suggested a morphometric neighbourhood, i.e. a neighbourhood relation that is weighted with the area of the Voronoi facet associated with the bond:
\begin{equation}
q_l^{(i)}=\sqrt{\frac{4\pi}{2l+1}\sum_{m=-l}^l\left| \sum_{f\in \mathcal{F}(i)} \frac{A_f}{A} Y_l^m[\theta_f, \phi_f]\right|^2}.
\label{eq:boo2}
\end{equation}
Here $A_f$ is the area of the facet $f$ associated with neighbour $j$, $A$ is the total surface area of the Voronoi cell, and the sum is over the ensemble of facets of the Voronoi cell $\mathcal{F}(i)$ of site $i$. In this definition the $q_l$ are unambiguous, parameter-free and continuous. Infinitesimal positional changes will create or annihilate facets with small areas and consequently the new or vanished bonds will only negligibly contribute to $q_l$.

The original formula for the $q_l$ (Eq.~\ref{eq:boo}), with using a Delauney-triangulation to determine neighbourhood, and eq.~\ref{eq:boo2}, using the morphometric neighbourhood definition with the additional facet-weighting, can be written into the general formula 
\begin{equation}
q_l^{(i)}= \sqrt{\frac{4\pi}{2l+1} \sum_{m=-l}^l\left| \sum_{f\in \mathcal{F}(i)} \frac{A_f^{\alpha}}{A(\alpha)} Y_l^m[\theta_f, \phi_f]\right|^2}.
\label{eq:booalpha}
\end{equation}
The exponent $\alpha$ of the facet area equals 0 for the Delaunay triangulation, which is just normalized by the number of bonds, and is 1 for the morphometric facet-weighting. The normalization $A(\alpha)$ of each bond reads $$A(\alpha)=\sum_{f\in \mathcal{F}(i)} A_f^{\alpha}.$$ The impact of the neighbourhood definition on the value of $q_l$ can be nicely illustrated using this parametrization. We use the $q_4$ and $q_6$ bond orientational order parameters. These parameters are most sensitive to the cubic and hexagonal symmetry, respectively, as we investigate here and are commonly used in studies on crystallization \cite{allahyarov_crystallization_2015,gasser_real-space_2001,rein_ten_wolde_numerical_1996,volkov_molecular_2002}. The $q_l$ of a lattice site in the ideal crystalline lattices are summarized in Tab.~\ref{tab:ql} and Fig.~\ref{fig:qlideal}. We compare the $q_4$- and $q_6$-values obtained using a distance threshold $r_c$ of 1.1 times the lattice constant $d$, and using the Voronoi-tesselation-based methods with an area-weighting exponent $\alpha$ of 0 an 1.

The $q_l$-values of the FCC and HCP lattice sites do not depend on the method of defining and weighting neighbourhood since they share only equal facets with the 12 nearest neighbours. The BCC lattice sites, in contrast, change their $q_l$-values by up to an order of magnitude with the different neighbourhood definitions, and the $q_4$ is more affected than the $q_6$. The strong dependence of the $q_l$ of the BCC lattice sites on the neighbourhood definition can be rationalized with their Voronoi cell (see inset in Fig.~\ref{fig:qlideal}). The Voronoi cells of the BCC lattice share facets with the 8 nearest, but also smaller facets with the 6 next-nearest neighbours. The 6 next-nearest neighbours are not counted as bonds when using a threshold, are counted equally when using Delauney- ($\alpha$=0)-weighting and are counted to less extent with morphometric $\alpha$=1-weighting.

We suggest that even better results can be obtained by using an $\alpha$=2 facet-weighting of the bonds. This is motivated primarily by the behaviour of $q_4$ when considering the BCC lattice. When a threshold is used such that only the 8 nearest neighbours are considered, a high fourfold symmetry compared to HCP and FCC is found (see Tab.~\ref{tab:ql}). If instead the next nearest neighbours are also taken into account and weighted equally, such as for $\alpha$=0, the fourfold symmetry is nearly completely destroyed. Heuristically, the more the close neighbourhood is emphasized, the easier it is to distinguish BCC by its fourfold symmetry from FCC and HCP. An $\alpha$=2 facet-weighting emphasizes the contribution from the larger fa\-cets assigned to the nearest neighbours compared to the facets of the next-nearest neighbours, thus is  closer to using a threshold than $\alpha$=1 facet-weighting. Still the $q_l$ stay a continuous measure as with the $\alpha$=1-weighting. The values of $q_4$ and $q_6$ of the ideal lattice sites are given in Fig.~\ref{fig:qlideal} and Tab.~\ref{tab:ql}. The difference among the $q_4$ and the $q_6$ of the ideal FCC and the ideal BCC lattice is larger using the $\alpha$=2-weighting than using the $\alpha$=1-weighting, which facilitates structure identification even in noisy crystals.

\section{\label{sec:resultsI} Application to noisy crystals}

\begin{figure*}
\centering
\subfloat{
\includegraphics[width=0.31\linewidth]{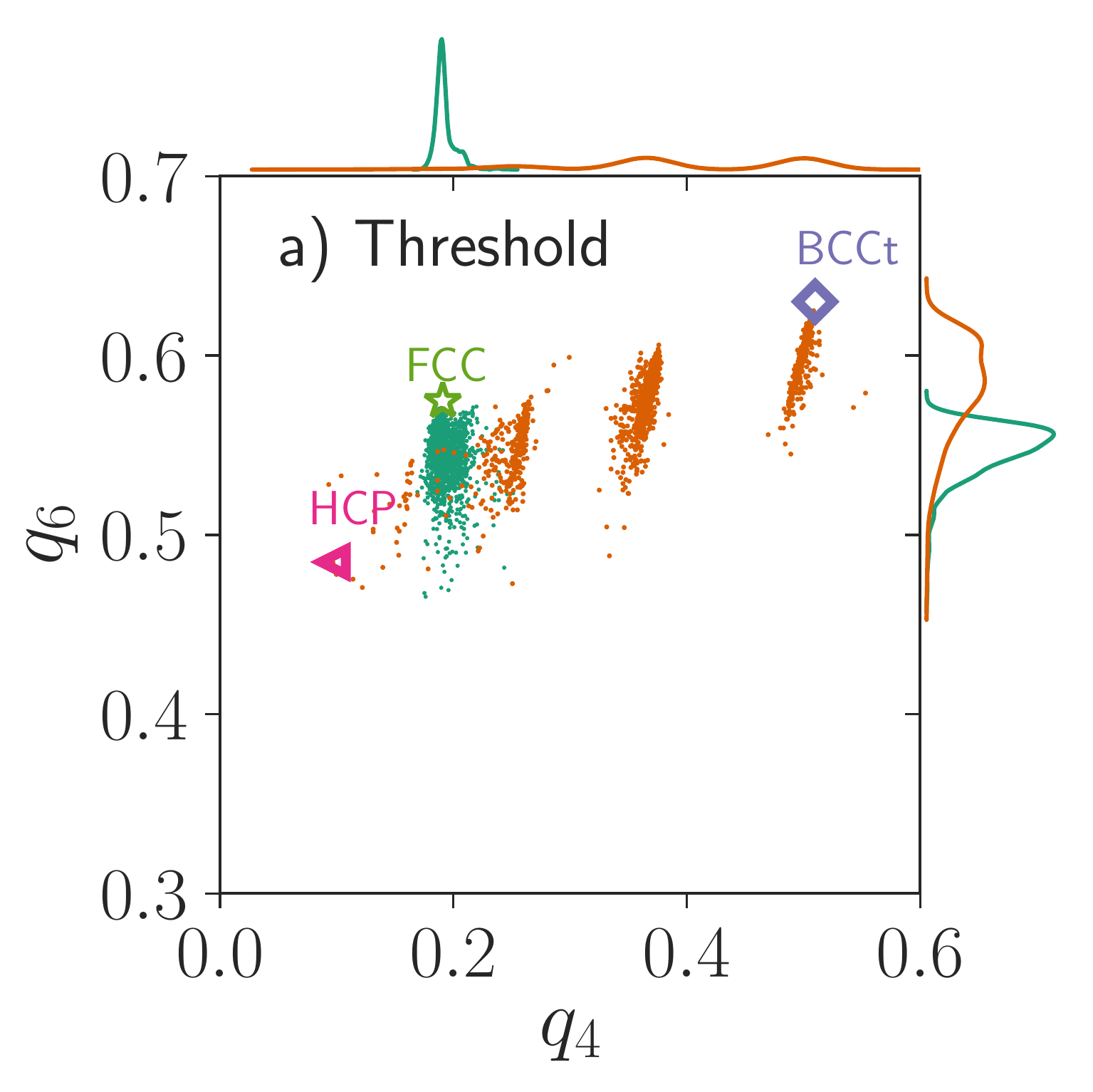}
}
\subfloat{
\includegraphics[width=0.31\linewidth]{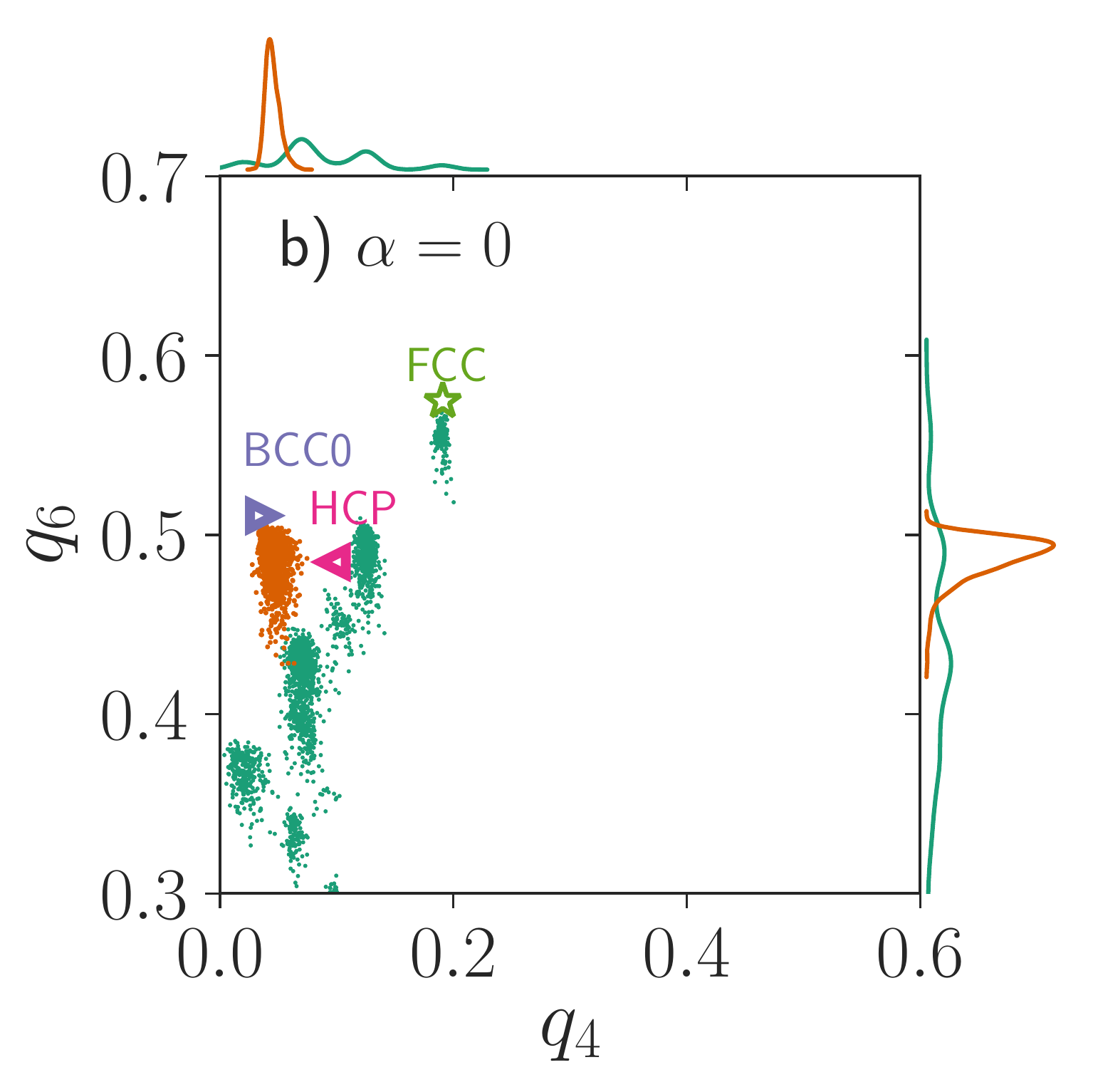}
}
\subfloat{
\includegraphics[width=0.31\linewidth]{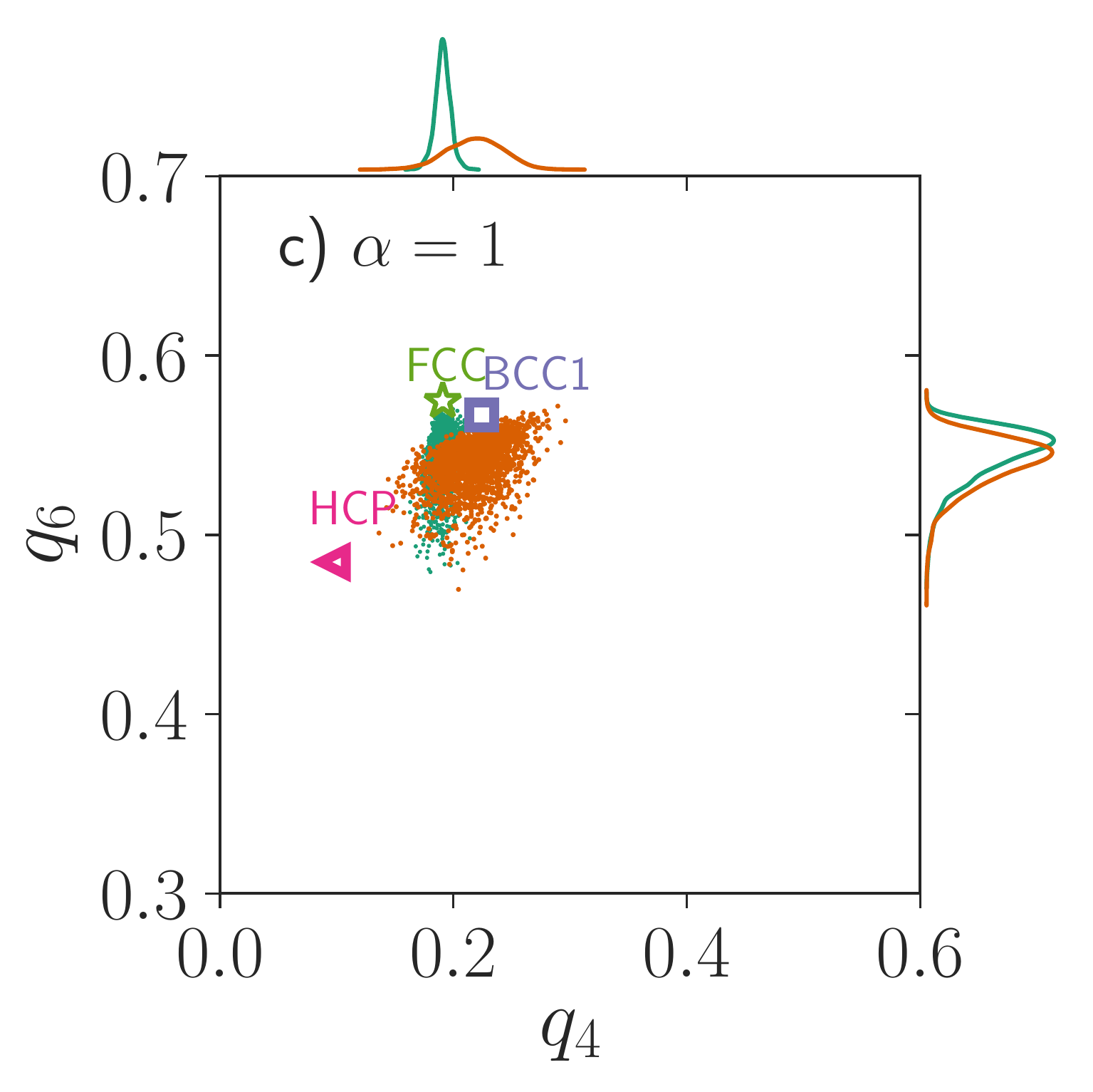}
}
\caption{Bond orientational order parameters of a BCC-based structure (orange) and a FCC-based structure (green) with a noise of $\sigma=0.036d$. In a) the neighbours are determined by a threshold value, in b) a Delauney tringulation is used ($\alpha$=0-weighting), in c) the Delauney neighbours are additionally facet area-weighted ($\alpha$=1-weighting). The graphs on the x- and y-axis give the distributions of the $q_4$- and the $q_6$-parameter individually on a linear scale. The large symbols indicate the $q_l$ values of the ideal lattices (see Tab.~\ref{tab:ql}).}
\label{fig:qlnoisy}
\end{figure*}

The dependence of the $q_l$-values on the weighting of the neighbourhood may be considered a nuisance in treatment of ideal structures. The numeric value of the $q_l$ shift, but still stay unique for the different crystal structures. This changes when noise is present in the structures.

Noise has severe impact on the $q_l$-values of the sites. We exemplify the effect of the noise when using the three established neighbourhood metrics in Fig.~\ref{fig:qlnoisy}. None of the three neighbourhood metrics can be considered satisfactorily. A splitting of the $q_4$-$q_6$-values of the individual lattice sites of the noisy BCC-structure into isolated clouds can be observed in Fig.~\ref{fig:qlnoisy}, a), where the threshold-definition of neighbourhood is used. The individual distributions of $q_4$ and $q_6$ on the axes of the graph show that the clouds of the BCC-structure already start to overlap with the cloud of the FCC-structure at this moderate noise of 3.6\%. The $q_4$-$q_6$-values of the noisy FCC-structure splits up into individual clouds in a similar fashion when using a Delauney triangulation to determine neighbourhood (Fig.~\ref{fig:qlnoisy}, b)). These clouds readily overlap with the values of the ideal HCP lattice and the cloud of the noisy BCC-structure for this moderate noise. Thus the $q_4$- and the $q_6$-values in a) and b) give the impression that the sampled structures consist of individual sub-populations with distinct local structures, in contrast to the homogeneous creation protocol and the continuously distributed displacements used to mimic the noise.

The reason for this behaviour is the discretized neigh\-bour\-hood-discrimination. With noise, a neighbour may be just below or above the threshold distance to be counted as a neighbour and the $q_l$-value jumps between distinct values (Fig.~\ref{fig:qlnoisy}, a)). Similarly, in Fig.~\ref{fig:qlnoisy}, b), minute displacements of the lattice sites leads to creation or annihilation of facets, and the $q_l$-value again jumps between distinct values. The BCC structure is particularly sensitive to the first mechanism because of the close distance between the nearest neighbours ($r=1d$) and the second nearest neighbours ($r=2/\sqrt{3}d\approx1.15d$). Any threshold chosen between those values will inevitably lead to misidentified neighbours once the first and second peak of the radial distribution function overlap (see Fig.~\ref{fig:gr},a)). The FCC structure is more sensitive to the second mechanism because its Voronoi cell contains six vertices which split and form new facets upon infinitesimal displacements as described by Troadec \emph{et al.} \cite{troadec_statistics_1998}. 

The $\alpha$=1 facet-weighting of the neighbourhood is continuous, contrary to the first two definitions (Fig.~\ref{fig:qlnoisy} c)). Emergence of a new facet with displacement of a lattice site creates a new neighbour with zero weight, and vanishing facets continuously contribute less to the $q_l$-value. However, even with this neighbourhood-weighting correct characterization of noisy crystal structures is difficult. The noisy BCC and FCC structures are constructed to be different, as can be seen from the radial distribution function (Fig.~\ref{fig:gr}). But this structural difference cannot be extracted from the $q_4$- and $q_6$-values, as the distributions of both parameters overlap.

\begin{figure}
    \centering
    \includegraphics[width=\linewidth]{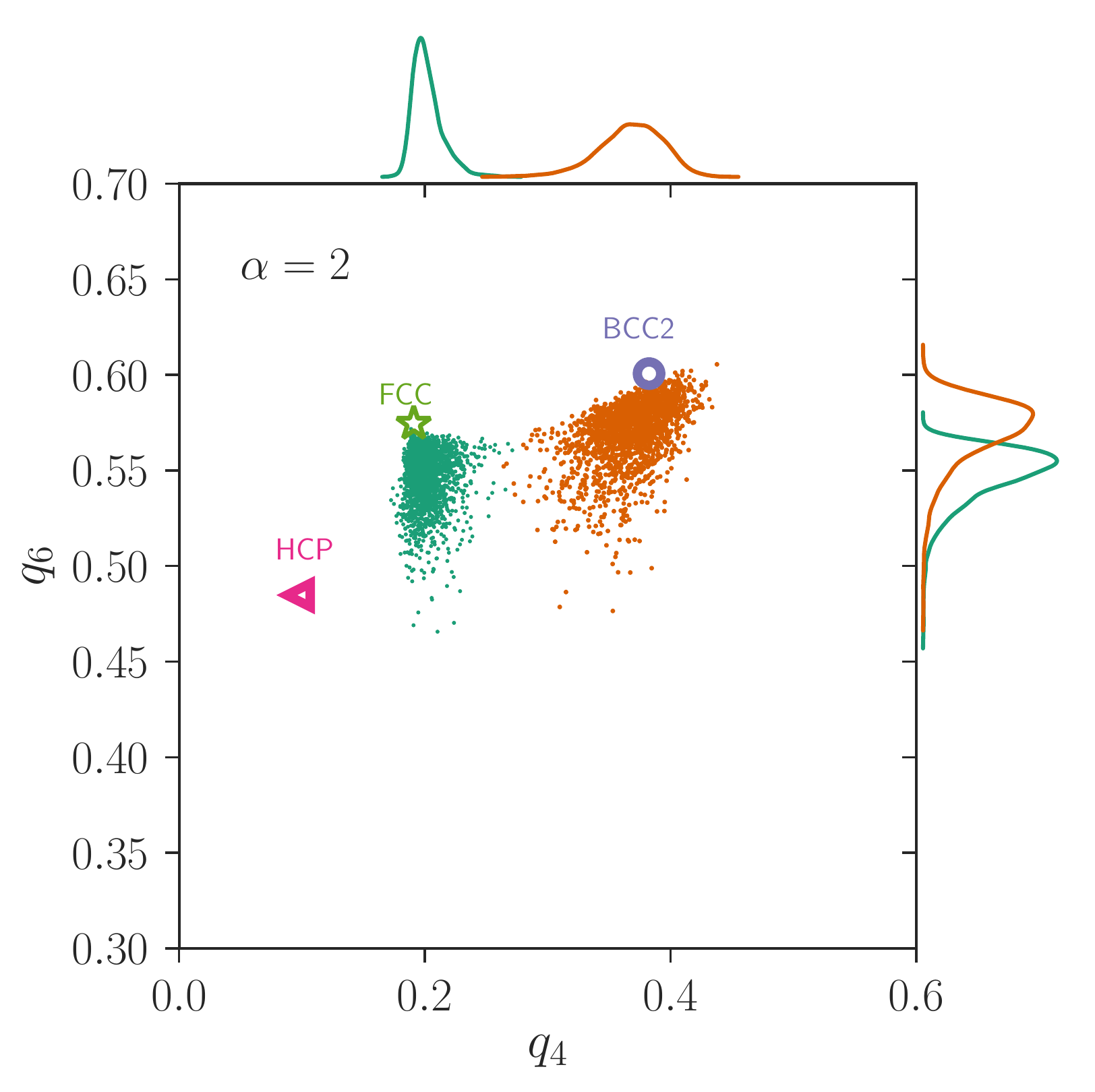}
    \caption{Bond orientational order parameter analysis of the noisy crystal structures using the $\alpha$=2 facet-weighted neighbourhood definition suggested in this paper. The $q_l$-values from the same noisy FCC and BCC structures as in Fig.~\ref{fig:qlnoisy} now form two continuous and distinguishable clouds of data points. The large symbols indicate the $q_l$ values of the ideal lattice sites (see Tab.~\ref{tab:ql}).
    }
    \label{fig:bondorder_vorosq}
\end{figure}

The result of a $q_4$-$q_6$-analysis using the $\alpha$=2 facet-weighted neighbourhood definition is displayed in Fig.~\ref{fig:bondorder_vorosq}. The analysed sites of FCC and BCC structures with noise generate two continuous clouds of points in the $q_4$-$q_6$-plane. This behaviour correctly displays the underlying homogeneous structures. In addition, the clouds of the BCC- and the FCC-structure are now clearly distinguishable, in contrast to the other neighbourhood definitions. Such a $q_4$-$q_6$-analysis consequently would allow for the analysis of crystallinity just by analyzing the distance of the noisy $q_l$ values from the values obtained for the ideal lattices.

\begin{figure*}
\centering
\subfloat{
\includegraphics[width=0.45\linewidth]{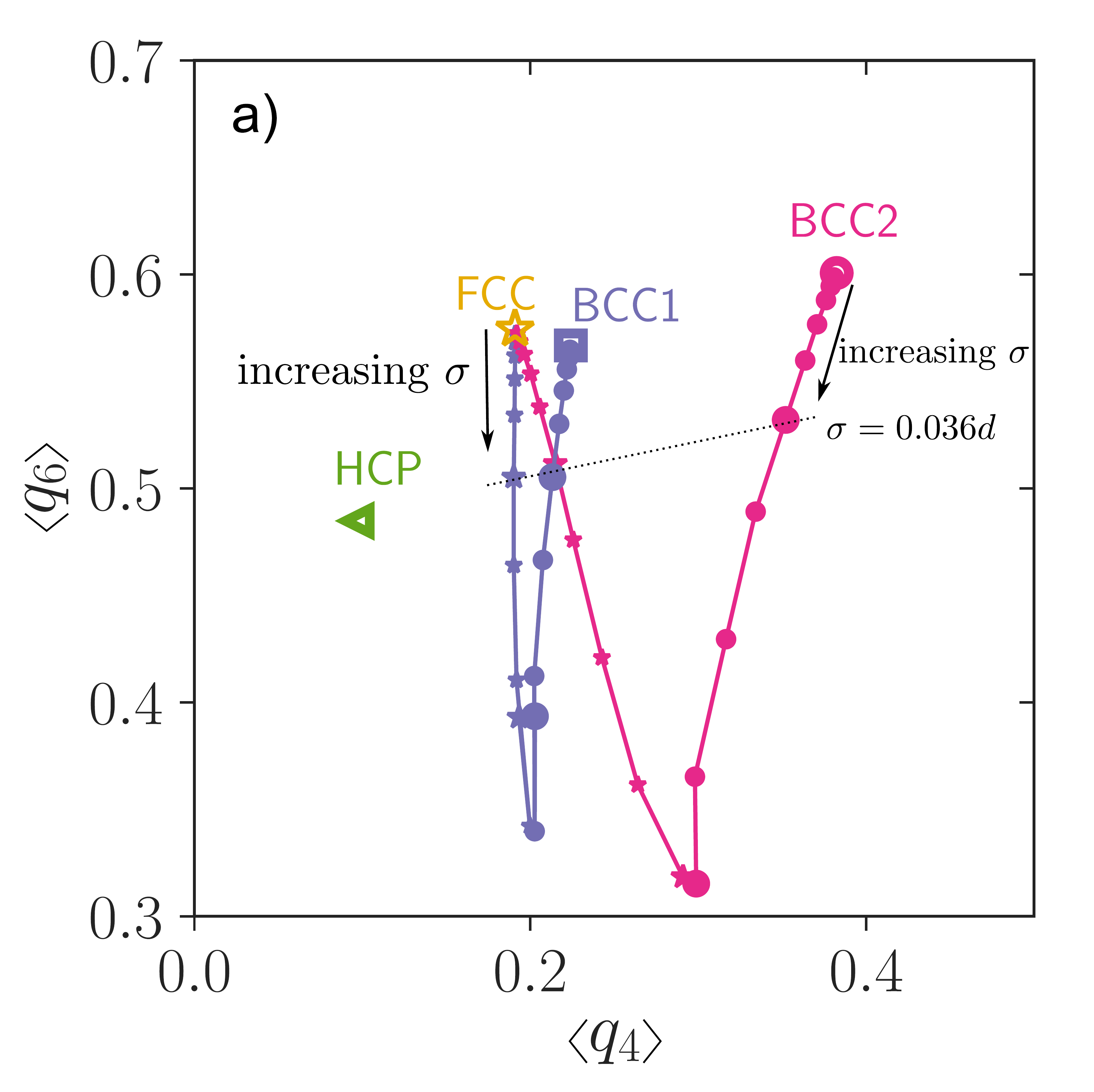}
}
\subfloat{
\includegraphics[width=0.45\linewidth]{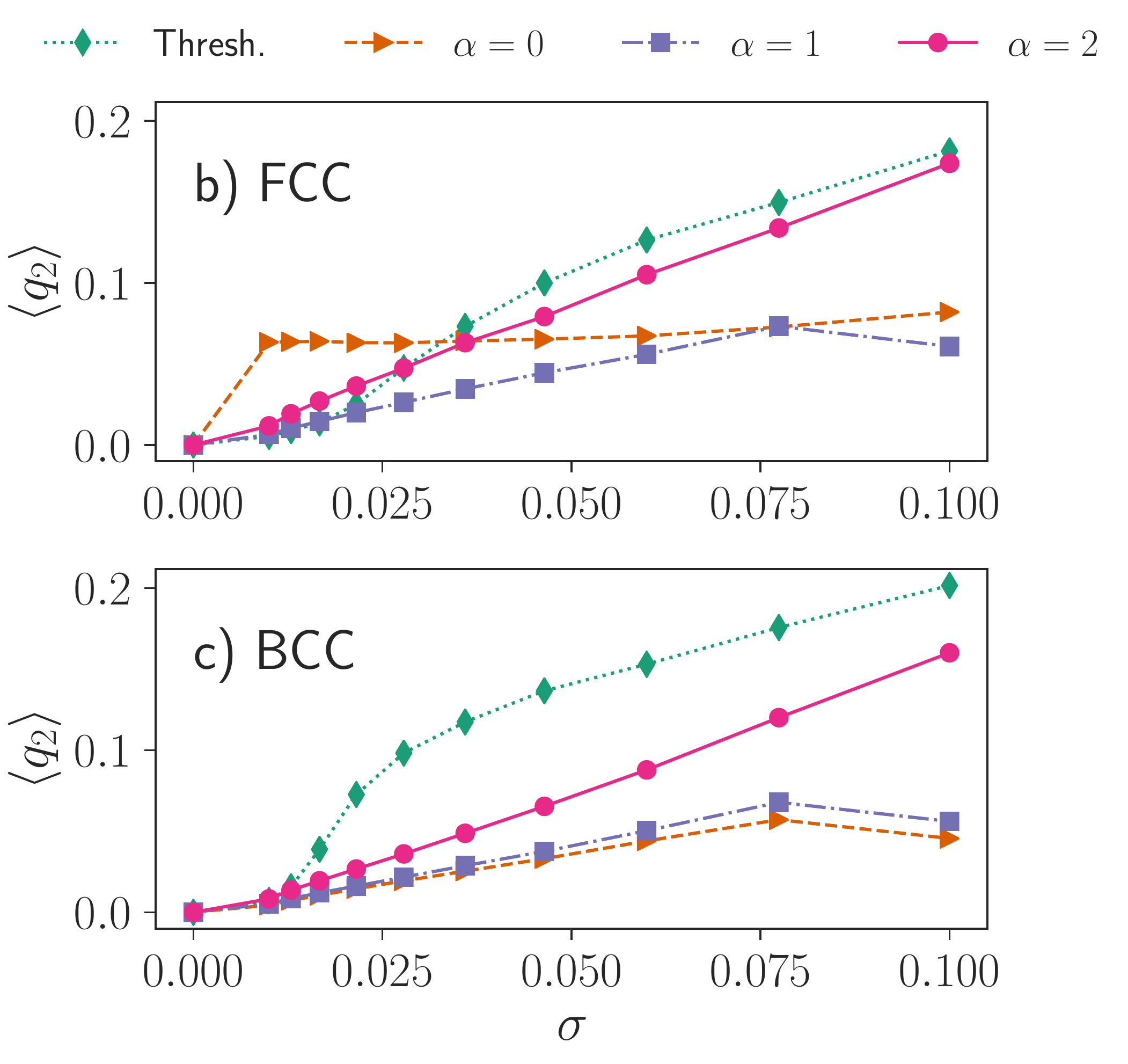}
}
\caption{Evolution of the average bond orientational order parameter attained from BCC- and FCC-based structures as a function of applied noise. In a) the evolution of the average $\left<q_4\right>$ and $\left<q_6\right>$ can be followed. The colour indicates the neighborhood definition (purple: $\alpha$=1, pink: $\alpha$=2) and the symbols indicate the noisy crystalline structure. From the values of the ideal, noise-free lattices (large symbols, compare Tab.~\ref{tab:ql}) the $q_6$-values continuously decrease, while the $q_4$ of the two crystal structures become increasingly similar. The $\alpha$=2-weighting emphasizes the differences in symmetry compared to the $\alpha$=1-weighting, and the values obtained for the two noisy crystal structures differ more. The dotted line indicates the $\sigma = 0.036$-noise level discussed in the figures above.\\ The evolution of the $\left<q_2\right>$ bond orientational order parameter with noise is displayed for the different neighbourhood definitions in the FCC-based structure, b), and the BCC-based structure, c). $\left<q_2\right>$  vanishes for both ideal, noise-free crystal lattices, but linearity and monotony of the growth of $\left<q_2\right>$ with noise depend on the neighbourhood definition.}
\label{fig:q2}
\end{figure*}

The optimization of sensitivity of the $q_4$-$q_6$-analysis using an $\alpha$=2-weighting can be seen for the whole tested range of noise in Figure~\ref{fig:q2}. The mean value of all $q_4$ and $q_6$ values obtained from the structures at a given noise level are displayed. The distance among the $q_4$-$q_6$-value of the FCC-based structure and the BCC-based structure at each noise level changes continuously with noise and becomes minimal with the largest noise values tested. The distance among $q_4$-$q_6$-positions of the FCC- and the BCC-based structure, which indicates the distinguishability of the two structures, is larger for the $\alpha$=2-weighting (BCC2) compared to $\alpha$=1-weighting (BCC1). Interestingly, the noise mainly affects and lowers the $q_6$-value, while the neighbourhood definition mainly affected the $q_4$-value of the BCC lattice sites (see Fig.~\ref{fig:qlideal}).

The analysis of structures is not limited to $q_4$ and $q_6$. These are commonly used, as they are especially sensitive to cubic or hexagonal symmetry, respectively.  Mickel \textit{et al.} suggested that the $q_2$ bond orientational order parameter could be of particular interest in studies on crystallinity. The $q_2$ parameter vanishes by any kind of mirror symmetry of the Voronoi cell \cite{mickel_shortcomings_2013}. It is thus not as useful as the $q_4$ and the $q_6$ in distinguishing between different crystalline structures. However, a vanishing $q_2$ serves as a good indicator of the presence of crystallinity.

We illustrate this behaviour in Fig.~\ref{fig:q2}, where the change of average $q_2$ with noise is displayed. $\langle q_2\rangle$ grows with increasing noise for all the neighbourhood definitions and vanishes for vanishing noise. The growth of $\langle q_2\rangle$ with noise is not continuous and monotonous in general, but with the $\alpha$=2-weighting. Care has to be taken especially with applying the $\alpha$=0-weighting neighbourhood definition, i.e. Delaunay triangulation, to the FCC structure. The $\langle q_2\rangle$ parameter does not approach 0 continuously as this metric exhibits a discontinuous jump as soon as infinitesimal noise is present (see discussion above and in Troadec \emph{et al.}\cite{troadec_statistics_1998}).

\section{\label{sec:conclusion} Discussion and Conclusion}

The Steinhardt bond orientational order parameters $q_l$ are a resourceful way to quantify geometric arrangement on a local scale. The comparability among studies is presently limited by the ambiguity how neighbourhood is defined. We have studied different definitions and weightings of neighbourhood in analyzing noisy crystal structures and suggested an optimization. None of the commonly used neighbourhood definitions would allow for a discrimination of a noisy BCC and a noisy FCC structure, as was highlighted in Fig.~\ref{fig:qlnoisy}. An arbitrary threshold and the Delauney triangulation ($\alpha$=0) introduce a discretized neighbourhood, and consequently the investigated homogeneous structures appear to bear different distinguishable local structures. The morphometric or $\alpha$=1 facet-weighting of neighbourhood already allows for an unambiguous definition of neighbourhood and continuous values of the $q_l$. However, this definition weights next-nearest neighbours in non-densest structures too heavily, consequently the symmetry of non-densest structures appears closer to the symmetry of densest structures. 

This drawback of the $\alpha$=1 facet-weighting is removed to large extent by using an $\alpha$=2 facet-weighting. The larger facets associated with the nearest neighbours are emphasized. By this non-densest structures can be distinguished from densest structures, an improvement not limited to BCC structures. In consequence, noisy crystalline structures can be more easily identified by mapping regions in a $q_4$-$q_6$ diagram to the ideal values (see Fig.~\ref{fig:bondorder_vorosq}). 

Application of $\alpha > 2$ weighting is also be possible. The neighbourhood metric stays unambiguous, continuous and emphasizes the nearest neighbours. However, we did not find a measurable improvement regarding the ability to distinguish noisy FCC and BCC structures for higher $\alpha$ values.

The $q_2$ bond orientational order parameter is useful for distinguishing amorphous and crystalline structures as it vanishes for a crystalline lattice. We show that the growth of $\left<q_2\right>$ with applied noise is continuous and monotonous for the suggested $\alpha$=2 facet-weighting.

\section*{Acknowledgements}
We thank Michael Klatt and Matthias Schr\"{o}ter for helpful discussions and Till Kranz for proof-reading the manuscript.

\section*{Author contribution statement}
JH has designed and analyzed the simulations. PB and JH have prepared the manuscript. MS has supervised the study and approved the final manuscript.

\bibliographystyle{unsrt}
\bibliography{minbib}

\end{document}